\documentstyle[11pt,twoside,psfig]{article}
\begin{document}

\begin{center}\LARGE\bf{Spectropolarimetry and the} \end{center}
\vspace{-24pt}
\begin{center}\LARGE\bf{Geometry of Type 1 Seyfert Nuclei}\end{center}

\begin{center}\Large{Andy Robinson, David J. Axon, James E. Smith} \end{center}

\begin{center}\normalsize{Department of Physical Sciences, University of Hertfordshire, College Lane, Hatfield, Hertfordshire, AL10 AB UK} \end{center}

\begin{center}\normalsize\it{To appear in Star Formation Through Time, ASP Conference Series} \end{center}
\vspace{-26pt}
\begin{center}\normalsize\it{eds: Per\'{e}z, Gonz\'{a}lez Delgado, Tenorio-Tagle} \end{center}


\begin{abstract}
We describe the results of a detailed study of the polarization 
properties of the broad H$\alpha$ line in Type 1 Seyfert nuclei.  Our 
analysis of these data points to a model in which the broad Balmer 
lines are emitted by a rotating disk, and are scattered in two main 
regions -- one co-planar with the disk and within the circum-nuclear 
torus, the other, the polar scattering region, outside the torus but aligned 
with its axis. The relative importance of the two sources of polarized 
light is largely determined by the inclination of the system 
axis to the line-of-sight.  
\end{abstract}

\section{Introduction}

In the Unified Model for Seyfert galaxies, Type 1 (S1)
and Type 2 (S2) Seyfert nuclei are the same type of object seen at
different orientations, our direct line-of-sight to the nuclear continuum
source and broad-line region (BLR) being blocked in the S2's by a
circum-nuclear torus of dusty molecular gas (e.g., Antonucci 1993).
Spectropolarimetry has played an important role in establishing this
picture; the detection of polarized broad-lines, attributed to scattering
of broad-line emission above the poles of the torus (e.g., Antonucci \&
Miller 1985), having revealed an obscured BLR in many S2's. 

Since the ${\bf E}$ vector for scattered light is oriented 
perpendicular to the scattering plane (the plane containing the 
incident and scattered ray) the polarization position angle (PA) will 
be perpendicular to the axis of the radio source, provided that the 
latter is co-aligned with the torus axis and hence the scattering 
cone.  This basic polar scattering picture broadly explains the 
optical polarization properties of S2 nuclei, in which the polarization 
PA is almost always perpendicular to the projected radio source axis 
(e.g., Antonucci 1983; Brindle et al.  1990).

In S1's, by contrast, the polarization ${\bf E}$ vector tends to be 
{\em aligned} with the radio source axis (Antonucci 1983, 1984; Martel 
1996).  Evidently, scattered light emerging from S1's follows a 
different path to that in S2's, implying that the simplest unification 
model geometry including only the polar scattering `mirror', is 
incomplete.  In particular, it seems that we require an additional 
source of scattered light in S1's to explain the alignment of the 
${\bf E}$ vector with the radio axis.  Our study of the optical 
polarization properties of S1's has led us not only to a better 
understanding of their scattering geometry but has also 
provided an important insight into the structure 
of the BLR.

\section{Observations and Results}

Our analysis is based on spectropolarimetry of the broad H$\alpha$ line 
in 36 Type 1 Seyfert, obtained during a number of different runs at the 
Anglo-Australian and William Herschel Telescopes.  The observations 
and results are described in detail in Smith 2002, Smith et al.  
2002a. The sample as a whole displays a wide diversity in 
polarization properties but, excluding objects which are significantly 
contaminated by foreground interstellar polarization, we can identify 
three broad categories.

Six objects have very low measured 
polarizations (consistent with null polarization) and are, therefore, 
likely be intrinsically weakly polarized, regardless of any 
interstellar contamination. 

The 20 intrinsically polarized objects exhibit a wide range of 
properties (Fig. 1). The average polarization is typically $\sim 1$\% for both 
the continuum and broad Balmer lines, but ranges from $\sim 0.5$ to 
$\sim 5$\%. Most of these objects exhibit significant variations in 
the degree ($p$) and/or position angle ($\theta$) of polarization 
across the broad H$\alpha$ line profile. The detailed polarization structure 
varies considerably from object to object but it is possible to discern 
certain common characteristics: (i) a blue--red swing in 
position angle across the H$\alpha$ profile and (ii) a central depolarization 
in the core of the profile, flanked by polarization peaks in the wings.

A few objects exhibit quite different polarization spectra.  These are 
characterised by a systematic increase in $p$ towards the blue end of 
the spectrum, with local increases associated with the broad H$\alpha$ 
and H$\beta$ lines.  Furthermore, in contrast to most of the 
intrinsically polarized S1's, there are no significant 
variations in $\theta$ over the broad-lines and indeed the PA 
is constant, to within a few degrees, over the entire observed 
spectral range.  These characteristics are remarkably similar to those 
of S2 nuclei in which polarized broad-lines have been 
detected.  We have firmly identified 2 such objects in our sample, 
Fairall 51 and NGC 4593, and more tentatively, a third, Was 45.  A 
literature search has revealed two more cases, Mrk 704 and Mrk 376 
(Goodrich \& Miller 1994).

Of the 10 objects for which the PA of the radio source axis is available,
the average position angles of polarization are approximately parallel to
the radio axis in 6 objects and approximately perpendicular in 3.

\begin{figure}
\psfig{figure=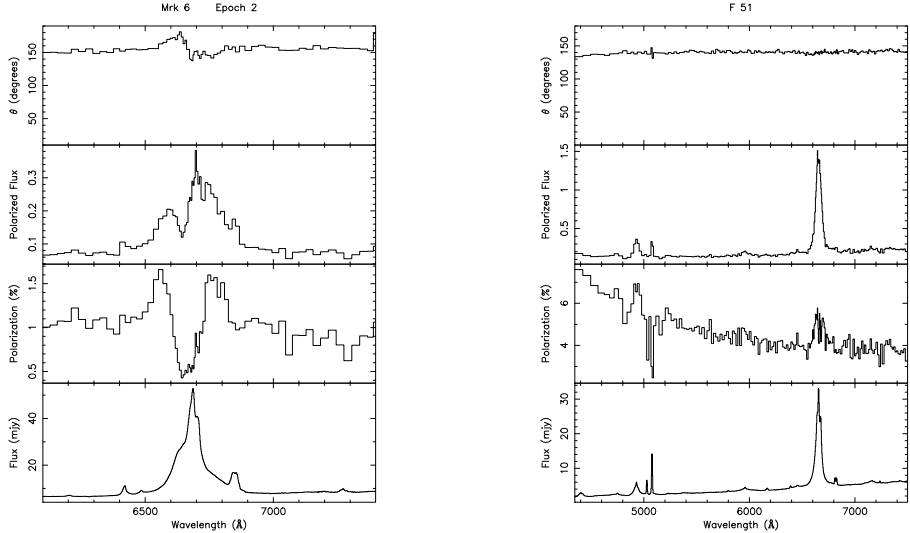,width=4.8in}
\centering{\caption{Polarization spectra of Mrk\,6 (left), an example of an equatorial scattering dominated S1 and Fairall\,51 (right), a polar-scattering dominated object.}}
\end{figure}

\section{Equatorial Scattering}

Among the objects whose average polarization PA is parallel to the 
radio axis is Mrk\,6, which also has the most striking variations in 
$p$ and $\theta$ across the broad H$\alpha$ profile (Fig.  1).  The 
orientation (averaged over wavelength) of the polarized light {\bf E} 
vector relative to the radio source indicates that scattering takes 
place in a plane perpendicular to the radio axis, i.e., in the 
equatorial plane of the torus.  The blue--red rotation in $\theta$ and 
the characteristic variation in $p$ are both naturally explained if 
the H$\alpha$ photons are emitted in a {\em rotating disk} and 
scattered in a co-planar ring closely surrounding the disk (Fig.  2).

The rotation in $\theta$ arises because red- and blue-shifted rays 
from opposite limbs of the disk subtend different scattering angles at 
each point in the ring.  When the disk is viewed face-on, circular 
symmetry ensures complete cancellation of the polarization produced by 
any point in the ring by that of its orthogonal counterpart.  However, 
when the disk is inclined, light scattered from points aligned with 
the projected minor axis of the disk is less completely polarized, due 
to the smaller scattering angle, than light scattered by points 
aligned with the projected major axis.  This breaks the symmetry and 
leaves a net polarization with a PA rotation similar to that for 
points aligned with the major axis.  The variation in $p$ is due to 
the fact that the scattering ring, being co-planar with the emitting 
disk, `sees' a broader profile than the observerÕs direct 
line-of-sight (which, in an S1, must necessarily be fairly 
close to the disk/torus axis).  The combination of scattered and 
direct line emission, therefore, results in wavelength-dependent 
dilution of the polarized component, producing the characteristic 
peak-trough-peak structure in $p$ across the line profile.  This 
`equatorial scattering' model is discussed in more detail by Smith et 
al.  2002b.

Apart from Mrk\,6 itself, 9 other objects in our sample display H$\alpha$ 
polarization properties consistent with the model.

\begin{figure}
\psfig{figure=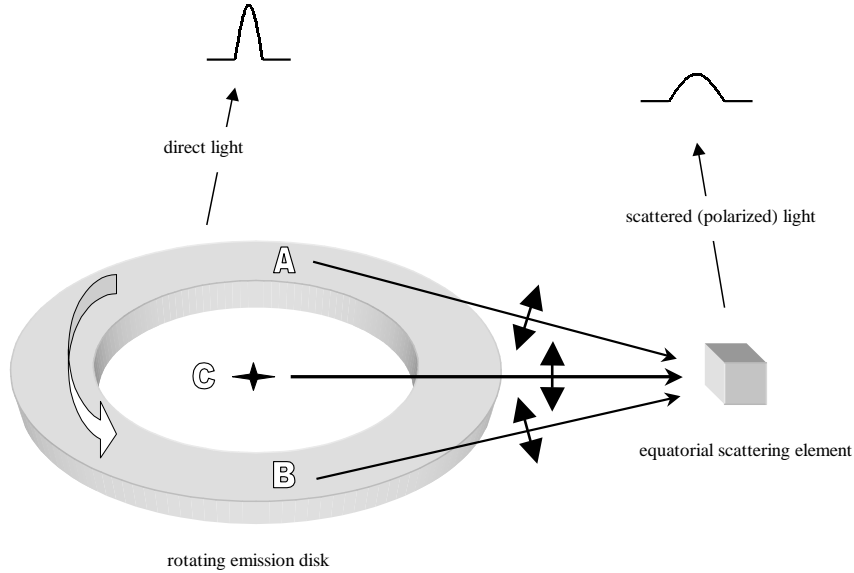,width=4.8in}
\centering{\caption{Equatorial scattering of H$\alpha$ emission from a rotating disk.}}
\end{figure}

\subsection{Implications for the Broad-Line Region}

It has frequently been postulated that the low ionization broad-lines 
(particularly the Balmer lines) in AGN are emitted by a rotating disk, 
presumably the outer regions of the accretion disk (e.g., 
Collin-Souffrin 1987; Murray \& Chiang 1997).  However, the observational 
evidence has, hitherto, been ambiguous at best.  We believe that our 
broad H$\alpha$ polarization data represents the most compelling 
evidence yet that a significant fraction of the line emission does 
indeed originate in a disk.  Although it is possible to conceive of 
other mechanisms that may explain either the PA swing or the variation 
in $p$ across the H$\alpha$ broad-line profile, the only plausible 
emission source--scattering region geometry that can simultaneously 
account for {\em both} of these properties is the model outlined 
above.  Moreover, since the amplitude of the $\theta$ rotation and 
precise form of the $p$ variation are sensitive to the 
disk--scattering ring distance, the radial extent of the disk and its 
inclination, detailed modelling of polarization properties and total 
flux profiles of the H$\alpha$ and H$\beta$ lines offers a unique and 
potentially powerful way of probing the structure of the disk and its 
environment.

\section{Polar Scattering}

Some S1's have polarization properties that are most readily explained 
by polar, not equatorial, scattering.  These include the cases in 
which the polarization PA is perpendicular to the radio source axis 
and in particular, those objects, like Fairall 51 (Fig.  1), which 
have polarization spectra similar to those of S2 nuclei with detected 
polarized broad-lines.

The polar-scattering region is visible in both Type 1 and 2 Seyferts 
since it is located outside the circum-nuclear torus and polarized 
flux from this region will therefore contribute to the net 
polarization in all Seyfert nuclei.  Polar-scattering clearly 
dominates in S2's but our results suggest that when the 
direct view of the nuclear regions is not blocked by the torus, 
equatorial scattering usually dominates the observed polarization.  
Why is this not the case in the `polar-scattered' S1's?

In the context of the Unified Model, an appealing possibility is that 
they are objects in which the inclination of the system axis is such 
that our line-of-sight to the nucleus passes through the relatively 
tenuous upper layers of the torus and is subject to a moderate amount 
of extinction; enough to suppress polarized light from the equatorial 
plane of the torus, but not the broad wings of the Balmer lines.  
Thus, the polarization spectrum is effectively that of an S2,
while the total light spectrum remains that of an S1.  We 
estimate that a visual extinction $A_V\sim 1$ mag along the 
line-of-sight to the equatorial scattering region would be sufficient 
to allow polar scattering to dominate the observed polarization (Smith 
et al. 2002c).  Interestingly, nuclear reddening estimates based on 
broad-band colours (Winkler et al.  1997) suggest that polar-scattered 
objects like Fairall 51, have somewhat higher nuclear extinctions than 
most S1's.

\section{Unification of Seyfert Polarization Properties}

The `polar-scattered' S1's provide direct evidence that the scattering 
cone situated above the torus, which is responsible for the polarized 
broad-lines seen in many S2's, is also present in S1's.  In our 
interpretation, the polar-scattered S1's represent a transition 
state between unobscured (the majority of Type 1) and obscured (Type 
2) Seyferts.  It follows that {\em all} Seyfert nuclei have both 
equatorial and polar scattering regions located, respectively, inside 
and outside the torus.  The latter is the scattering cone, which is 
co-axial with the torus (and radio) axis.  The former is co-planar 
with, and closely surrounds, a rotating disk that emits a significant 
fraction of the broad Balmer lines.

Four inclination regimes, which produce quite distinct polarization 
signatures can be identified (Smith et al. 2002c; also  
http://star-www.herts.ac.uk/\\$^{\sim}$jsmith/sequence2.htm).  
When the system is viewed 
almost face-on ($i\approx 0^\circ$), both the equatorial and polar 
scattering regions exhibit a high degree of circular symmetry and 
cancellation leads to null or very low polarization.  At intermediate 
inclinations ($0 << i < 45^\circ$), there is no extinction along the 
direct line-of-sight to the nucleus and both scattering regions, as 
well as the broad-line region, are visible.  In general, equatorial 
scattering dominates the observed polarization.  When the inclination 
of the system axis is comparable to the torus opening angle ($i\approx 
45^\circ$) the line-of-sight to the nucleus is subject to a moderate 
amount of extinction and polar-scattered S1Õs are observed.  At still 
larger inclinations ($i>45^\circ$), both the BLR and equatorial 
scattering region are completely obscured by the torus and the 
broad-lines are only visible in polarized light scattered from the 
polar scattering region.  A Seyfert Type 2 with polarized broad-lines 
is observed.

The range of polarization properties exhibited by Seyfert galaxies can,
therefore, be broadly understood in terms of an orientation sequence based
on the two-component scattering model.

\section*{References}

\begin{flushleft}
Antonucci R.R.J. 1983, Nature, 303, 158\\
Antonucci R.R.J. 1984, ApJ, 278, 499\\
Antonucci R.R.J. 1993, ARA\&A, 31, 473\\
Antonucci R.R.J. Miller J.S., 1985, ApJ, 297, 621\\
Brindle C., et al. 1990, MNRAS, 244, 577\\
Collin-Souffrin, S. 1987, A\&A, 179, 60\\
Goodrich R.W., Miller J.S. 1994, ApJ, 434, 82\\
Martel A.R., 1996 PhD Thesis, UCO/Lick Obs., Univ. 
California, Santa Cruz\\
Murray N., Chiang J. 1997, ApJ, 474, 91\\
Smith, J.E., 2002 PhD Thesis, Univ. Hertfordshire\\
Smith, J.E., et al. 2002a, MNRAS, 335, 773\\
Smith, J.E., et al. 2002b, MNRAS, submitted\\
Smith, J.E., et al., 2002c,MNRAS, submitted\\
Winkler H. 1997, MNRAS, 292, 273\\
\end{flushleft}

\section*{Discussion}

\noindent {\it Tim Heckman and Brian Boyle:} Is the geometrical and 
kinematic structure you deduce from the BLR consistent with the picture 
that is infered from reverberation mapping? \\

\noindent {\it Mark Whittle:} In addition to establishing the presence 
and geometry of the near and far field scattering medium, do these 
observations also demand a rotating disk geometry for the BLR, and if 
so, is that consistent with the reverberation results?  \\

\noindent {\it David Axon:} In answer to both questions, as far as we 
can see, the only plausible geometry for the emission source that 
naturally explains, in combination with equatorial scattering, {\em 
both} the rotation in $\theta$ and the variation in $p$ across the 
broad H$\alpha$ line is that of a rotating disk.  As to whether this 
structure is consistent with reverberation mapping results, we would 
first say that while the latter generally seem to favour a 
gravitationally-dominated BLR, the precise structure (e.g.  whether a 
disk or spherical cloud ensemble) is not strongly constrained.  
However, we are not claiming that {\em all} of the broad H$\alpha$ 
emission comes from the disk -- if, for example, a virialized cloud 
ensemble co-exists with a disk, the former would only contribute a 
constant polarization vector (because there is no velocity 
discrimination), which would not affect the {\em form} of the $\theta$ 
variation across the profile.  In effect, spectropolarimetry is only 
sensitive to the disk component.\\

\end{document}